\documentclass[12pt]{article}
\let\section=\subsection     \let\subsection=\subsubsection                
\usepackage{graphicx}

\begin{document}
\begin{center}
   {\large \bf Short versus long range interactions and}\\[2mm]
   {\large \bf the size of two-body weakly bound objects}\\[5mm]
   R.J. Lombard$^{a)}$ and C. Volpe$^{a),b)}$ \\[5mm]
   {\small \it $^{a)}$ Groupe de Physique Th\'{e}orique, 
Institut de Physique Nucl\'{e}aire, \\
   F-91406 Orsay Cedex, France \\
$^{b)}$ Institut f\"ur Theoretische Physik der Universit\"at, \\
    Philosophenweg 19, D-69120 Heidelberg, Germany \\[8mm]}
\end{center}

\begin{abstract}\noindent
Very weakly bound systems may manifest intriguing ``universal'' properties,
independent of the specific interaction which keeps the system bound.
An interesting example is given by relations between the size
of the system and the separation energy, or scaling laws. So far,
scaling laws have been investigated for short-range and long-range (repulsive)
potentials. We report here on scaling laws for weakly bound two-body systems
valid for a larger class of
potentials, i.e. short-range potentials having a repulsive core and
long-range attractive potentials. We emphasize analogies and differences
between the short- and the long-range case. In particular, we show that
the emergence of halos is a threshold phenomenon which can arise 
when the system is bound not only by short-range interactions but also 
by long-range ones, and this for any value of the orbital angular momentum
$\ell$ ! 
These results enlarge the image of halo systems we are accustomed to.

\end{abstract}

\section{Introduction}

Weakly bound systems have attracted
a lot of attention in different domain of physics. 
Halo nuclei have been studied for several years 
using various probes \cite{KR}. 
Diffuse Van-der-Waals dimers and trimers represent 
another interesting example, such as $(^{4}He)_2$  whose
existence has been established only recently \cite{DD,Lew,euro}.
All these systems are characterized by very large spatial extensions
on one hand and very small separation energies on the other hand.
For example, the one-neutron halo $^{11}Be$ has a 
separation energy of about 0.5 MeV, the
average separation energy being of 6 MeV, while the average distance between
the halo neutron and the core is about 7 fm.
The $(^{4}He)_2$ dimer is a paragon:
its predicted dissociation energy is 0.1 $\mu$eV, 
whereas the ionization energy is of about 27 eV and
the measured average distance between the
two atoms is of about 52 $\AA$ \cite{DD,st}. This is by far 
the largest dimer ever observed ! 

One of the intriguing features of very weakly bound systems
is that they can reveal ``universal'' behaviors, i.e. independent of the
specific interaction which keeps together the particles.
An example is given by
the relation between the size of the system, which can be
characterized through
different moments of the wave function, like the mean square radius
or $\langle r^2 \rangle$, and the (separation or dissociation)
energy $E_{S}$ necessary to break the
system, that is  
\begin{equation}\label{e:1}
\langle r^2 \rangle = f(E_{S})~~~~E_{S} \rightarrow 0.
\end{equation}
This relation is often called (asymptotic) scaling law.

Several works exist on the topic of scaling laws
for two- and three-body systems obtained with
short-range and 
repulsive (mainly Coulomb) long-range 
interactions \cite{past,euro,R}. For the short-range case,
so far only potentials defined by two parameters,
the depth and the range, have been considered.

Recently these investigations have been widened by the study
of a larger class of potentials  \cite{LV}, 
namely : {\it i)} short range potentials having
a repulsive core 
to simulate the Pauli exclusion principle;
 {\it ii)} long-range attractive potentials.
Here we report on this investigation and describe how scaling
laws for short-range potentials are modified by the presence
of the core. We present scaling laws for long-range interactions and
emphasize analogies and differences between
the short- and the long-range case.

Finally, 
we come to a question which has been
extensively discussed in the literature \cite{euro,past}, 
namely : 
What is a halo system ?
The image we are now accustomed to is the one in which  
the halo particle has a very large probability of being outside the
classically allowed region. 
This image needs to be enlarged if one wants to include 
the case of long-range potentials, as we will discuss.

\section{Theoretical Framework}

Scaling laws for two-body systems (\ref{e:1})  can be treated in the context of the 
two-body problem in quantum 
mechanics\footnote{In the case of halo nuclei, such a description
has a validity as far as the
degrees of freedom of the halo nucleons 
can be separated from those of the core.}. A bound state of the system
is then identified as a bound state of the 
potential 
$V(r)$\footnote{Here we will consider that the system has spherical symmetry so that
the potential is a function of the relative distance between
the two bodies only. A dependence on the spin degrees of freedom does not change our
conclusions.}
 acting 
between the two atoms or the core and the halo nucleon in a nucleus.
The separation energy of the system $E_S$ is taken equal to the energy of the
bound state $E_{\ell}$.

There are (at least) three possible ways to tackle this problem \cite{LV} :
\begin{itemize}
\item[*]
Using very general arguments, without specifying the potential $V(r)$.
These arguments are based, for example, on inequalities like the Bertlmann-Martin
inequality, or on properties of the Schr\"odinger 
equation\footnote{Since we consider only the lowest states of each angular momentum (the
wave function has no node), they are simply labeled by $\ell$.}
 ($\hbar = 2m = 1$) : 
\begin{equation}\label{e:2}
[ - \frac{\partial ^2}{\partial r^2} - \frac{2}{r} \frac{\partial}{\partial r}
+ \frac{\ell(\ell+1)}{r^2} + \lambda V(r) ] 
\psi_{\ell}(r)= E_{\ell}\psi_{\ell}(r) , 
\end{equation}
such as its invariance under
the 
transformation\footnote{This transformation
is valid for potentials defined by two parameters only.} 
$x=r/R_0$ and $\epsilon=E_{\ell}R_0^2$, where $R_0$ characterizes the
range of the potential.

\item[*]
Once the range of the potential is defined,  
in the limit of very small binding one can make 
the following approximations : 
{\it i)} for short-range potentials the wave function is mainly given by the tail outside 
the potential, i.e. by
the Hankel wave function
$\psi_{\ell}(r)  \approx  e^{-\mu r}/r^{\ell + 1}$ 
with $\mu(E_{\ell})$;
 {\it ii)} for long-range potentials the wave function, mainly inside the potential, can be taken 
as the
spherical Bessel function of the
first kind 
$\psi_{\ell}(r)  \approx j_{\ell}(kr)$ with $k(E_{\ell})$,  cutting the integrals 
at the first zero. 
\item[*] Specific potentials can be used to get quantitative estimates.

\end{itemize}

\section{What happens to Scaling Laws for Short-Range Potentials having
a Repulsive Core ?}

For the class of short-range potentials going to zero
faster than $1/r^{2}$,  
the scaling laws are \cite{euro,past} :

\begin{equation}\label{e:3}
\langle r^2 \rangle_0 \approx \frac{c_0}{|E_0|} \ ,~~~~~ \ 
\langle r^2 \rangle_1 \approx \frac{c_1 R_0}{\sqrt{|E_1|}} \ ,~~~~~ \
\langle r^2 \rangle_2 \approx {c_2 R_0^2}  \ ,
\end{equation}
for s- ($\ell=0$), p- ($\ell=1$) and 
d-states\footnote{There is no energy dependence in the scaling laws of states
having $\ell>2$ as well.} 
($\ell=2$).
As we can see, the mean square radius diverges as $1/|E_0|$ and 
as $1/\sqrt{|E_1|}$ for s- and p-states respectively, but no divergence is present
for states having $\ell \ge 2$.
If we take the divergence of the second moment of the wave function as a reference
for the appearance of halos, we can say that halos may occur for s- and p-states, in the
limit of very weak binding, but that such phenomenon cannot appear for states of higher
$\ell$.

The relations (\ref{e:3}) have been obtained using the behavior of the tail of the
wave functions \cite{past}. This automatically implies that the  
constants $c_{\ell}$ are independent of the potential $V$.
Numerical studies have also been performed in the past 
to check the sensitivity of $c_{\ell}[V]$
to the specific shape of the potentials; but so far potentials defined by their depth and
their range 
(like the square well and the Gaussian potential) have been considered. 

However, strictly speaking the scaling law 
is completely independent of the shape of the short-range potential for the s-state only, 
since $c_0=1/2$ 
\cite{LKL}. 
Can physical systems, like halo nuclei or diffuse dimers,
attain this region of very small binding, where the scaling
law (\ref{e:3}) become completely independent of the potential ?
This can be easily checked by looking if 
$\langle r^2 \rangle_0{|E_0|}=1/2$.
We show results for $^{11}Be$\footnote{The same is true for the other halo 
nuclei \cite{R}.} and
the $(^{4}He)_2$ 
dimer\footnote{We use the $\langle r^2 \rangle$ from calculations in
\cite{Lew} which well reproduce the measured $\langle r \rangle$ of $(^{4}He)_2$.}, 
as examples (Fig.1).
We can see that if nuclear halo systems are located in a region which
is still sensitive to the specific choice of the potential, diffuse 
dimers attain the region of very weak binding \cite{R,LV}.

Concerning the other coefficients  $c_{\ell}$ ($\ell >0$), 
if one considers a larger class
of short-range potentials, defined by more than two-parameters, like those including a repulsive core, the independence of (\ref{e:3}) from the potential
is no more strictly true \cite{LV}. 
To illustrate how $c_{\ell}$ depend on $V$,
we compare the results obtained with a two parameters' potential and with a potential
including a repulsive core (Fig.1).

We can see that the higher $\ell$ is, the stronger is the dependence of $c_{\ell}$
and therefore of the corresponding scaling law (\ref{e:3}) on the shape of the potential.
In fact, the role played by 
the centrifugal barrier is the larger, the higher is $\ell$.
As a consequence, the wave function is pushed more and more inside the potential and becomes
sensitive to its specific shape.

\section{What are Scaling Laws for Long-Range Potentials?}
Using both inequalities and the behavior of the wave functions we have shown that 
the scaling laws for the class 
of long-range potentials 
going to zero slower than  $1/r^{2}$ (such as 
the attractive Coulomb interaction, the
confining potentials like the linear potential and the harmonic oscillator) 
are \cite{LV} :
\begin{equation}\label{e:4}
\langle r^2 \rangle_{\ell} \approx  \frac{c_{\ell}[V]}{|E_{\ell}|}~~~~~\forall \ell
~~~~~E_{\ell} \rightarrow 0 \ .
\end{equation}
For the s-states, the scaling law (\ref{e:4}) has the same dependence on
$1/|E_{0}|$ as for short-range potentials. However, the coefficients 
$c_{0}$ here (as well as the other $c_{\ell}$ with $\ell >0$)
 will always depend on the potential $V$ because the 
wave function is mainly inside the potential and feels its specific shape.
For example, $c_{0}=3$ for the Kratzer potential, the Coulomb attractive potential
and the harmonic oscillator. 

\begin{center}
   \includegraphics[width=9cm,height=8cm,angle=-90]{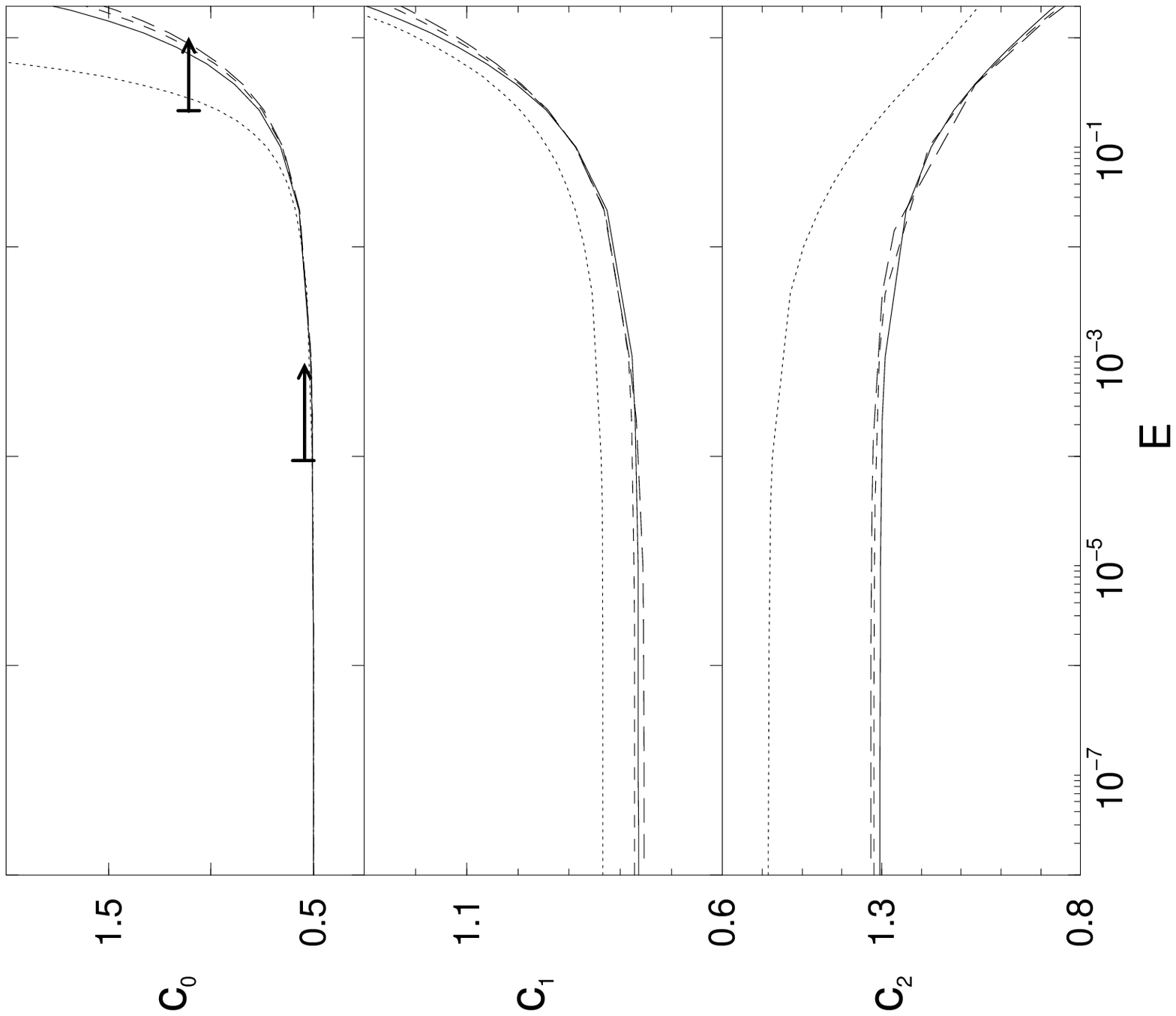}\\
   \parbox{14cm}
        {\footnotesize 
        Fig.~1: Scaling laws (\ref{e:3}) 
obtained with short-range potentials for s- (top), p- (middle)
and d-states (bottom) \cite{LV} :  The coefficients $c_{\ell}$  are shown as a function
of the energy $E_{\ell}$ of the state. The full, the short-dashed and long-dashed lines
correspond to a potential including a repulsive core with three different
ranges $R_0$. As an illustration the cut Kratzer potential is taken: 
$V_{cur}(r)=-2 ({a \over r}-{a^2 \over {2r^2}})\Theta(R_{0}-r)$.
For comparison the results obtained with a potential without a repulsive core
(the square well) are shown (dotted line). For $c_0$,  
the arrows indicate  the range of the 
energies corresponding to halo nuclei (right) and to diffuse
Van-der-Waals dimers (left).}
 \end{center}

For the states having $\ell > 0$, the mean square radius always diverges 
as $1/|E_{\ell}|$, contrary to the short-range case (\ref{e:3}).
Therefore systems kept together by a long-range interaction 
can develop very large extensions, and this for states of any angular momentum $\ell$ !
This difference with the known short-range case results can be intuitively
understood. In fact, in the long-range case the centrifugal barrier
is not playing a major role in confining the halo particle inside the
potential. 
 
An example of very extended 
systems bound by a long-range interaction is already known :
Rydberg atoms.
However, there is a major difference.
Here we are predicting that there may exist 
very weakly bound two-body systems
bound by long-range interactions having large spatial extensions
in their lowest energy states.

\section{What is a Halo System?}

How to define a halo system has been the object of extensive
discussions in the literature.
Nowadays we think of halo systems as systems in which the
halo particle has a very high probability of being outside
the classical turning point \cite{euro,past}.
In the case of long-range potentials the particle is mainly in
the classical allowed region. The appearance of halos can still be characterized 
in a common way for both the short- and the long-range cases : 
the $\langle r^2 \rangle_{\ell}^{1/2}$ has to be
much larger than a typical physical length of the system 
(such as for example the size of the core).
   
These new results tell us that the appearance of halos is a threshold
phenomenon, independent of the range, short or long, of the interaction 
and
therefore also of the time spent by the halo particle
inside or outside the classical region.

\end{document}